# Facile and rapid synthesis of highly luminescent nanoparticles via Pulsed Laser Ablation in Liquid

G Ledoux[1,2,3,6], D Amans[1,2,3], C Dujardin[1,2,3] and K Masenelli-Varlot[4,5]

[1]Université de Lyon, Lyon, F-69000, France,
[2]Université Lyon 1, Villeurbanne, F-69622, France. CNRS,
[3]UMR5620, Laboratoire de Physico-Chimie des Matériaux Luminescents, F-69622 Villeurbanne, France,
[4]Université de Lyon, INSA-Lyon, MATEIS,
[5]UMR CNRS 5510, F-69621 Villeurbanne cedex, France.

E-mail: ledoux@pcml.univ-lyon1.fr

**Abstract:** This paper demonstrates the usefulness of pulsed laser ablation in liquids as a fast screening synthesis method able to prepare even complex compositions at the nanoscale. Nanoparticles of $Y_2O_3$:$Eu^{3+}$, $Lu_2O_2S$: $Eu^{3+}$, $Gd_2SiO_5$:$Ce^{3+}$ and $Lu_3TaO_7$:$Gd^{3+}$,$Tb^{3+}$ are successfully synthesized by pulsed laser ablation in liquids. The phase and stoichiometries of the original materials are preserved while the sizes are reduced down to 5-10 nm. The optical properties of the materials are also preserved but show some small variations and some additional structures which are attributed to the specificities of the nanoscale (internal pressure, inhomogeneous broadening, surface states…).

Submitted to : Nanotechnology

## 1. Introduction

In the last decades, many publications have been devoted to luminescent studies of nanomaterials. For instance these luminescence properties at the nanoscale were first observed in II-VI compounds like CdS, CdSe… where the strong quantum confinement effect on their gap leads to the well-known pictures where the color changes according to their sizes.[1] The studies were then extended towards other semiconductors like Silicon[2,3] or ZnO.[4] More recently,[5] doped insulator nanomaterials have attracted more and more attention due to their higher chemical and radiation stability. For these materials, the luminescence properties are governed mainly by the activators and the confinement effects are at the second order level.[6, 7, 8] On the other hand, a lot of well-known ionic compositions are widely used as bulk in many applications such as lighting, display, lasers, scintillators. It is therefore attractive to transfer some specific optical properties from bulk to nanoscale compounds (e.g. biolabelling ) or to study the size effects on these properties. In this frame, many different synthesis methods have been developed over the years like co-precipitation methods, sol-gel methods, solvothermal processes, (see

---

[6] Author to whom any correspondence should be addressed

Cushing *et al.*[9] for a review of the different chemical route methods), combustion,[5] mechanosynthesis[10] or pyrolysis.[11] Each method requires optimization according to each composition, which is often time consuming for complex stoichiometries. We propose in this paper a fast screening synthesis method able to prepare even complex compositions at the nanoscale, with mono-dispersed distributions and preserved stoichiometries and phase.

Pulsed Laser Deposition (PLD) in vacuum or in a controlled atmosphere has been widely studied and is a very powerful technique to grow thin films of complex stoichiometries.[12] It is only since the end of the nineties that researchers have started to be interested in pulsed laser ablation in liquid media (PLAL). This method combines the advantages of standard PLD and those of soft chemical routes. As with the soft chemical routes, the product obtained is a stable colloidal dispersion of nanoparticles in a liquid medium but the ablation process allows the growth of materials with complex stoichiometries. Moreover, using pulsed laser ablation in a liquid medium gives access to materials which can only be synthesized at high pressure. Briefly, when a target is irradiated with fluences over 0.1 $GW.cm^{-2}$, material is ejected and evaporated. According to Fabbro *et al.*,[13] a laser power density of several $GW.cm^{-2}$ ensures a maximum pressure of several GPa generated by shockwaves. In the literature, direct measurements of the shockwaves properties generated on a silicon substrate in water[14] and on an aluminum foil in water (5 GPa)[15] have been reported. This effect allows for example the synthesis of diamond[16,17,18,19] and $BN$[20] phases.

Apart from high pressure phases, this method has shown the ability to make a great variety of nanocrystals including metal nanoparticles (Au, Ag, Pt, In, Fe, W, Cu and their alloys[21,22,23,24]), semiconductors nanocrystals (GaAs, CdSe, CdTe,[25,26] ZnO,[27]) and more recently some insulator nanocrystals ($TiO_2$, $SnO_2$,[28] $Eu_2O_3$, $CeO_2$, $LiCoO3$, $CeTbO_3$[29,30]). Another advantage of this method is the ability to dope the material up to very high concentrations by adding some additional elements in the solution. Indeed, $C_3N_4$ particles have been prepared by ablating a carbon target in an ammonia solution.[31] The size of the particles can be controlled by using surfactant molecules.[27,28] Moreover, since the particles are in a solution after synthesis, they can be functionalized for safe and simple manipulation or even used in biological media.

The use of this method opens up the possibility of studying new materials at the nanoscale range and therefore to envision new applications. Since PLAL was not used to synthesize materials for their luminescence properties, we focused our paper on some typical composition of luminescent materials. We tested several families of materials, sesquioxydes, oxysulfides, silicates and tantalates. We show for each family a complete study for a given stoichiometry, respectively $Y_2O_3:Eu^{3+}$, $Lu_2O_2S:Eu^{3+}$, $Gd_2SiO_5:Ce^{3+}$, $Lu_3TaO_7:Gd^{3+}, Tb^{3+}$.

Briefly, $Y_2O_3:Eu^{3+}$ is a well-known phosphor used in lighting and display applications. $Lu_2O_2S:Eu^{3+}$ and $Gd_2SiO_5:Ce^{3+}$ are famous scintillating materials. The oxysulfide family is well-known to be delicate to synthesize. In the case of $Gd_2SiO_5:Ce^{3+}$, the challenge consisted in the control of cerium valence since it is known that as bulk Cerium tends to be tetravalent. Lastly, the ultra dense material $Lu_3TaO_7:Gd^{3+},Tb^{3+}$ (d=9.49 $g.cm^{-3}$) was produced to test the co-doping in a complex composition.

## 2. Experimental Section

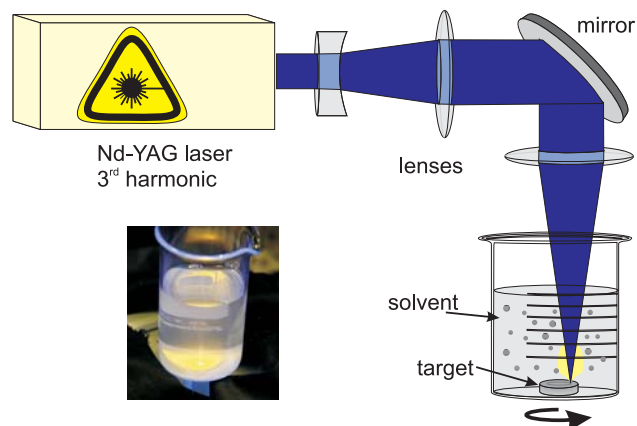

**Figure 1:** Scheme of the PLAL experiment. The laser is focused on the target which is rotated. A plasma is generated from which the nanoparticles are formed. On the picture, one can see the impact on the target as well as the plasma plume and its luminescence.

Facile and rapid synthesis of highly luminescent nanoparticles via Pulsed Laser Ablation in Liquid

Target for laser ablation were prepared either by standard solid state reaction techniques ($Y_2O_3$:$Eu^{3+}$, $Lu_3TaO_7$:$Gd^{3+}$, $Tb^{3+}$, $Gd_2SiO_5$:$Ce^{3+}$) or by pressing commercially available powders ($Lu_2O_2S$:$Eu^{3+}$). In the case of the Cerium doped sample the annealing steps were performed under an Ar/$H_2$ (5% $H_2$) mixture to ensure the trivalence of Cerium ions. A schematic of the PLAL experiment is shown on figure 1. The targets were placed at the bottom of a 50 ml beaker and covered by 40 ml of de-ionized water. A third harmonic YAG laser ($\lambda$=355 nm, $\Delta t$=5 ns, repetition rate 10Hz) was focused at the surface of the target. The laser illuminated the target until the solution became milky (Table 3 reports the duration of the experiment for each composition). The upper 30ml of solution was collected afterwards. Part of the solution was lyophilized in order to collect powders. The amount of material produced is quite small. Therefore to perform X-Ray diffraction the powder is deposited on a micro-lasso made of nylon and the measurement are performed on a Gemini S Ultra system from Oxford diffraction. The x-ray source used has a Molybdenium anode (Mo K$\alpha$1 at 0.7093Å) focusing the beam on a 300μm spot. Additionally statistics on transmission electron microscopy (TEM) images and analysis were performed. TEM measurements were performed on a JEOL 2010F microscope operating at 200 kV and equipped with Energy Dispersive X-ray Spectroscopy (EDS). For TEM measurements a few droplets of the solution were poured onto 300-mesh copper grids covered with a holey carbon film. Water was then eliminated by contact with filter paper. Images were recorded using a slow-scan CCD camera and the electronic diffractogram on all high resolution images were calculated with the Digital Micrograph Software. EDS analysis were performed on 5 different zones of the TEM grid at the lower possible resolution in order to average over a great number of particles. Photon correlation spectroscopy (PCS) was performed on a Zetasizer Nano ZS from Malvern instruments. The luminescence of the samples was checked from the solution and from the powder in a home made spectro-fluorimeter. The excitation is provided by a 450 W Xenon lamp whose discharge is focused at the entrance of a Gemini 180 monochromator from Jobin-Yvon. At the exit of the monochromator the light is refocused on the sample. The luminescence is then collected by an optical fiber and fed into a second monochromator (Triax 320 from Jobin Yvon) and detected by a photomultiplier (EMI 9789) or a Peltier cooled CCD. Excitation spectra were performed with the photomultiplier when the emission was below 600nm. The resolution in excitation used was 2 nm in this case ($Gd_2SiO_5$:$Ce^{3+}$, or $Lu_3TaO_7$:$Gd^{3+}$, $Tb^{3+}$). For Europium doped samples emission spectra were automatically recorded with the CCD while varying the excitation wavelength. The excitation spectra were reconstructed afterward from the resulting excitation/emission mapping. In this second case the resolution in excitation used was 2 nm for $Y_2O_3$:$Eu^{3+}$ and 1 nm for $Lu_2O_2S$:$Eu^{3+}$. For the emission spectra the resolution was 1 nm for $Gd_2SiO_5$:$Ce^{3+}$ and 0.25 nm for all the others. A resulting advantage of the complete emission/excitation mapping is the possibility to detect luminescence from impurities or unwanted phases. All excitation spectra were corrected from the wavelength dependence intensity of the lamp which was calibrated with a Powermeter 1931-C from Newport.

## 3. Results and discussion
*3.1. Structural characterization*

Facile and rapid synthesis of highly luminescent nanoparticles via Pulsed Laser Ablation in Liquid

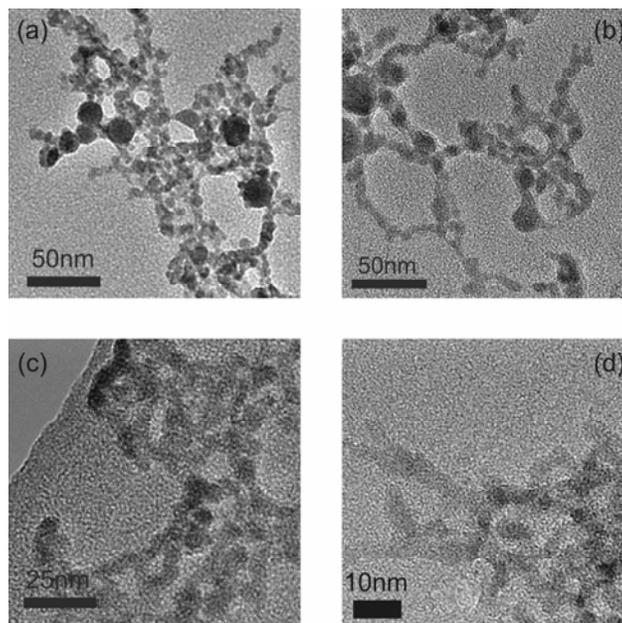

**Figure 2:** Transmission electron microscopy images of four different samples at low magnification: (a) $Lu_3TaO_7$:$Gd^{3+}$,$Tb^{3+}$, (b) $Y_2O_3$:$Eu^{3+}$, (c) $Gd_2SiO_5$:$Ce^{3+}$ and (d) $Lu_2O_2S$:$Eu^{3+}$

Conventional TEM images of the different samples are shown on figure 2. Particles form a web-like structure but the contrast is high enough so that the primary particles are easily distinguishable. For instance results from the TEM image analysis of $Gd_2SiO_5$ nanoparticles which were obtained using the procedure described by Vermogen *et al.*[32], are presented on figure 3. The results for all the samples are reported in Table 1. It appears that the mean sizes of the primary particles always lie in the range of 7 nm. The standard deviation is between 2 and 3 nm and only few particles have sizes lower than 3 nm or higher than 25 nm. The logarithm of the experimental distribution underwent a Henry test combined with a complete linear regression analysis (including the residues analysis), with a level of confidence of 95%. For each composition, the parameters of the corresponding log-normal distribution (mean value or average diameter, standard deviation), are summarized in Table 1 together with the 90% and 98% bilateral confidence intervals (computed with the previously determined log-normal distribution parameters).

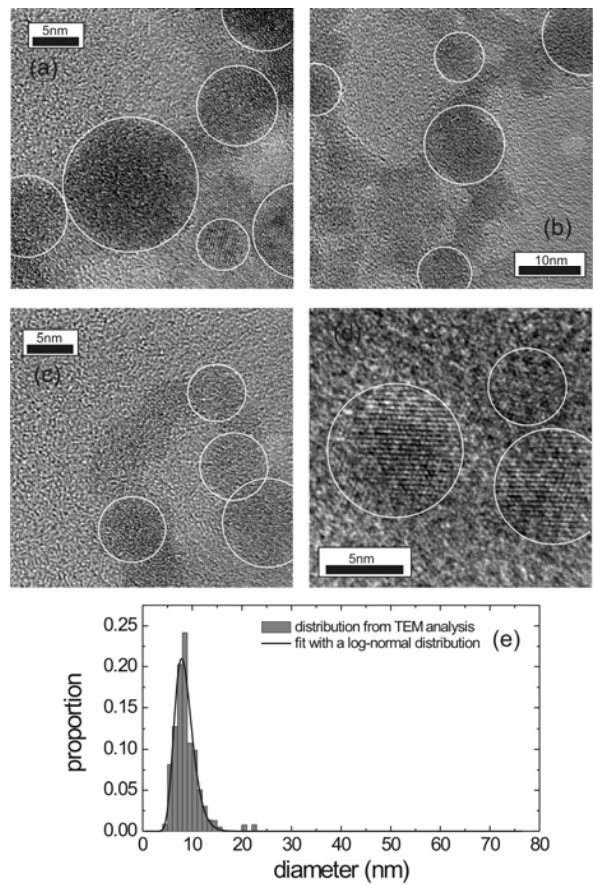

**Figure 3:** examples of image analysis used to get the size distribution in the case of $Gd_2SiO_5:Ce^{3+}$. The white circles are some guides to the eye giving the average diameter of individual particle. For higher resolution some particles (those which are well oriented with the electron beam) show diffracting planes (see image d) which helps in the determination of the limit of the particles. (e) Example of size distribution deduced from the analysis of the TEM images. The distribution is well fitted with a log normal distribution with an average size of 8.9nm.

The observed particles size distributions correspond to a log-normal distribution which is typical from accretion growth methods[33]. It means that the particles grow in the plume and are not directly extracted from the pellet (this last case would give rise to a $1/d^n$ size distribution)[34]. In water, the individual particles are then aggregated and form a web-like structure. The hydrodynamic size of the particles in solution determined by photon correlation spectroscopy (PCS) show that the sizes of this aggregates are in the range between 100 to 200nm in good agreement with the observations of the aggregates by electron microscopy. Figure 4 presents high resolution images of the individual particles.

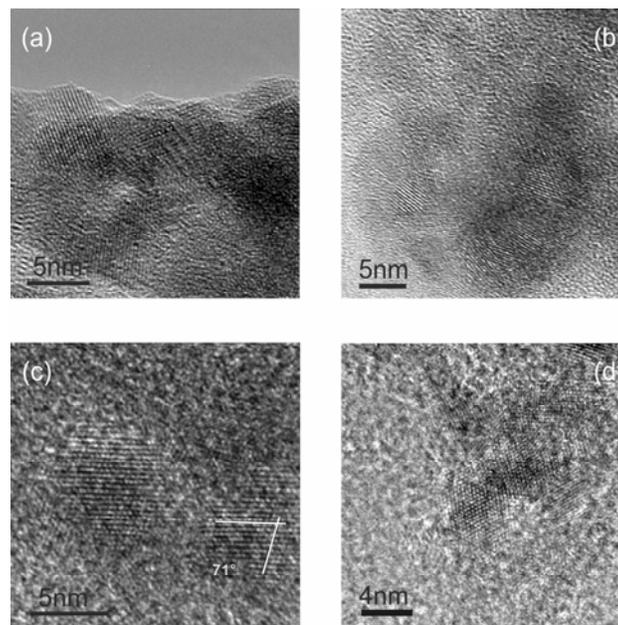

**Figure 4**: High resolution transmission electron microscopy images of 4 different samples: (a) $Lu_3TaO_7$:$Gd^{3+}$,$Tb^{3+}$, (b) $Y_2O_3$:$Eu^{3+}$, (c) $Gd_2SiO_5$:$Ce^{3+}$ and (d) $Lu_2O_2S$:$Eu^{3+}$. In the case of $Gd_2SiO_5$:$Ce^{3+}$ (Image (c)) one can clearly see a nanoparticle for which two families of planes are apparent and which form an angle of 71°.

Nanoparticles appear well crystallized and because of their high ionicity they appear slightly facetted even at very small sizes. The electronic diffractograms corresponding to the images of figure 4 are reported on figure 5. The diffracting planes of the different compositions are clearly observed.

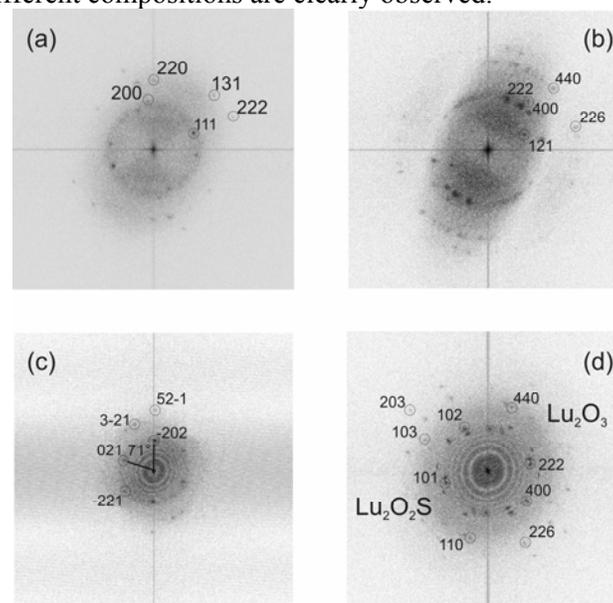

**Figure 5**: Electron diffractogram for the 4 different samples of figure 4: (a) $Lu_3TaO_7$:$Gd^{3+}$,$Tb^{3+}$, (b) $Y_2O_3$:$Eu^{3+}$, (c) $Gd_2SiO_5$:$Ce^{3+}$ and (d) $Lu_2O_2S$:$Eu^{3+}$. In each case the corresponding diffracting plane families are indicated. In the case of $Gd_2SiO_5$:$Ce^{3+}$ (image (c)) planes -202 and 021 correspond to the same particle and form an angle of 71° as expected. For $Lu_2O_2S$:$Eu^{3+}$ (image (d)) both $Lu_2O_3$ and $Lu_2O_2S$:$Eu^{3+}$ family planes are observed. Those of $Lu_2O_3$ are indicated in the right part of the diffractogram while those of $Lu_2O_2S$:$Eu^{3+}$ are indicated in the left part.

To check the phases obtained, the inter-planar distances were compared with their bulk counterpart data from the literature or from the JCPDS files for all the high resolution TEM images acquired. For $Lu_2O_2S$ we used the JCPDS file n°26-1445 and for $Lu_3TaO_7$ the JCPDS file n°24-0697. In the case of $Y_2O_3$, we used the JCPDS file n°25-1011. For $Gd_2SiO_5$ we used the data from Dramicanin *et al.*[35] so that we could get the angles between similar families of planes. For all the compositions we find a good agreement with the results for the bulk phases. Nevertheless $Lu_2O_2S$ shows additional planes corresponding to the cubic $Lu_2O_3$ phase. Energy dispersive X-Ray Spectroscopy (EDS) of the same images is presented in the supporting information and the quantitative analysis over 5 different zones observed at low resolution is reported on Table 2. For $Lu_3TaO_7$, we observe an excess of Tantalum although the electron diffractogram indicates that no evidence of pure tantalum or tantalum oxide are observed and that the phase is that of the bulk. In the case of $Gd_2SiO_5:Ce^{3+}$ the ratio of elements is well preserved. Finally, in the case of $Lu_2O_2S$ there is a strong excess of Lutetium, which is coherent with the presence of the diffracting planes of $Lu_2O_3$. From the EDS analysis $Lu_2O_3$ represents 60% of the particles and $Lu_2O_2S$ 40%. Due to the $Lu_2O_3$ phase contamination, no statistics on the size distribution are presented.

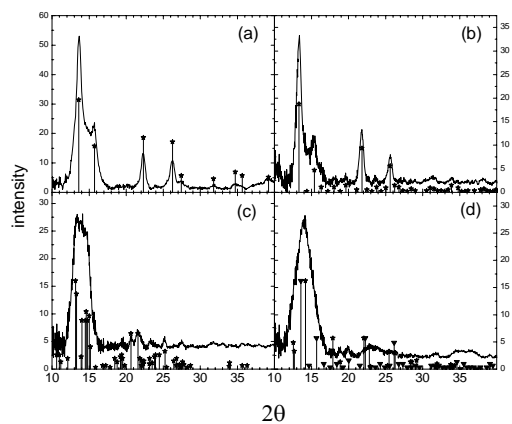

**Figure 6**: X-Ray diffraction for the 4 different samples: (a) $Lu_3TaO_7:Gd^{3+},Tb^{3+}$ compared with JCPDS file n°024-0697 of $Lu_3TaO_7$ (★); (b) $Y_2O_3:Eu^{3+}$ compared with JCPDS file n° JCPDS 025-1011 of $Y_{1.9}Eu_{0.1}O_3$ (★); (c) $Gd_2SiO_5:Ce^{3+}$ compared with JCPDS file n° 00-040-0287 of $Gd_2SiO_5$ (★) (d) $Lu_2O_2S:Eu^{3+}$ compared with JCPDS file n° 026-1445 of $Lu_2O_2S$ (★) and JCPDS file n° 012-0728 of $Lu_2O_3$.(▼)

Figure 6 presents the results of X-ray diffraction analysis performed on the different samples. The spectra confirm the observation obtained with the transmission electron microscope. The different phases are preserved in the cases of $Lu_3TaO_7$, $Y_2O_3$ and $Gd_2SiO_5$ while part of $Lu_2O_2S$ has been transformed into $Lu_2O_3$.

*3.2. Spectroscopic properties*

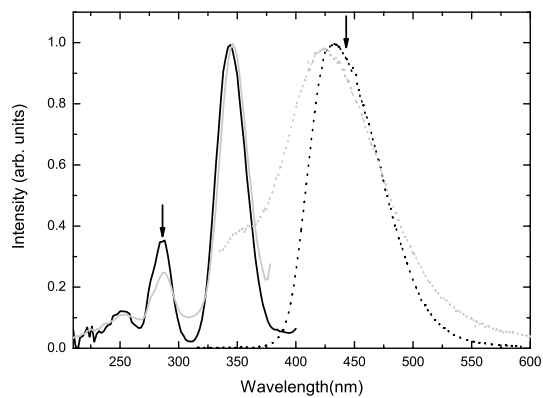

**Figure 7**: Emission (dashed lines) and excitation (continuous lines) luminescence spectra of bulk (black) and nanoparticles (grey) grown by PLAL of Cerium doped $Gd_2SiO_5$. For the excitation spectra the light was detected at 450nm, for the emission spectra the samples were excited at 290nm.

Here the optical properties of the nanoparticles synthesized by PLAL are compared to those of the targets. The main goal is to check if the main spectral properties are preserved and to use the luminescence of the rare earth as structural probes.

Figure 7 shows excitation and emission spectra for both samples, the bulk target of $Gd_2SiO_5:Ce^{3+}$ and the nanoparticles obtained by PLAL. For the excitation spectra the emitted light was detected at 450 nm (indicated by an arrow). For the emission spectra the samples were excited at 290 nm. The emission and excitation are fully explained by considering the transitions between the configurations of the $4f^1$ and $4f^05d^1$ states of $Ce^{3+}$. Due to its large radial extension, the 5d orbital interacts strongly with the matrix. The crystal field effects on the configuration levels of the $4f^05d^1$ induces a strong dependence of the emission and absorption wavelength as a function of the cerium surroundings.[36] Since the spectra from both samples (bulk and nano-particles) are roughly similar, the optical spectroscopy confirms that the nano-particles crystallize in the same phases and stoichiometries as the bulk ones. However from the emission spectra, two differences can be pointed out: (i) a slight blue shift (6nm) of the spectrum maximum, from bulk to nanoparticles, and (ii) a slight broadening of the spectrum for the nanocrystals. Both observations can be interpreted by considering the crystal field effects and slight modifications of cerium ions' surroundings when the crystal size is reduced to the nanometric scale. The inter-atomic distance can indeed depend on the nanoparticle size because of the Gibbs pressure.[37] Therefore, the shift is defined by the average crystal field effect due to the average lattice distortion, and so to the mean size of the nanoparticles, while a part of the broadening can be due to the standard deviation of the size distribution. Besides, in a single nanoparticle, a slight disorder appears at the nanometric scale.[6] This disorder also implies an inhomogeneous broadening of the spectrum.

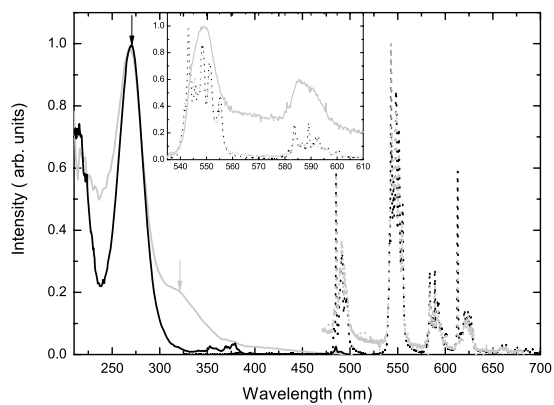

**Figure 8**: Emission (dashed lines) and excitation (continuous lines) luminescence spectra of bulk (black) and nanoparticles (grey) grown by PLAL of Terbium/Gadolinium doped $Lu_3TaO_7$. For excitation spectra the light was detected at 550nm, for emission spectra the samples were excited at 275nm except for the spectrum with a continuous grey line in the inset that was excited at 325nm.

Figure 8 shows excitation and emission spectra for the bulk target of $Lu_3TaO_7:Gd^{3+}$, $Tb^{3+}$ and the nanoparticles obtained by PLAL. For the excitation spectra the light was detected at 550nm. The emission spectra were obtained under 275 or 325 nm excitations. The emission observed between 480 nm and 700 nm corresponds to the well known $^5D_4$-$^7F_J$ transitions of $Tb^{3+}$, with J between 0 and 6. The $Gd^{3+}$ ions do not exhibit any emission in the visible range. However, their presence results in a shoulder, around 310 nm, in the excitation spectrum. This bump is due to $^8S_{7/2}\rightarrow{}^6P_J$ transitions of $Gd^{3+}$, while the wide band centered at 275 nm is due to the 4f-5d inter-configurational absorption of $Tb^{3+}$ ions. When going from bulk to nanoparticles, different changes can be observed. In the excitation spectra, the $4f^8$-$4f^75d^1$ absorption band is broadened for the same reason as reported before in the case of $Ce^{3+}$ in $Gd_2SiO_5$. Besides another broad band centered at 325 nm appears. In the case of the emission spectrum for this excitation wavelength (solid line in the inset) the spectrum is strongly modified. The different Stark components of the $^5D_4$-$^7F_J$ transitions are blurred and now form only broad peaks. This is typical of some sort of strong structural disorder in a material and is probably the signature of the smaller particles in the size distribution.

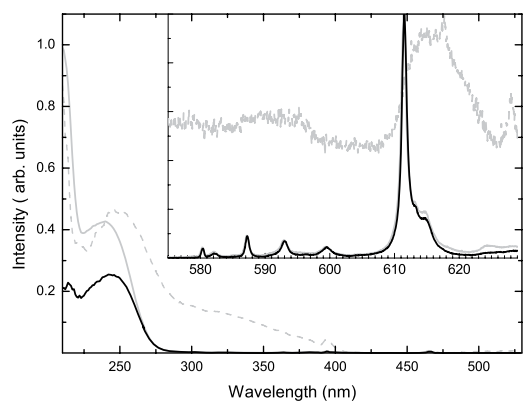

**Figure 9**: Excitation spectra of Europium doped yttrium oxide nanoparticles grown by PLAL (grey lines) for the emission at 611nm (continuous line) or at 620nm (dashed line) or in the bulk phase (black line). In the inset emission spectra of bulk (black line) excited at 250nm and of the nanoparticles excited at 250nm (continuous line) and 350nm (dashed line).

Figure 9 presents the spectroscopy of the Europium doped $Y_2O_3$ samples. The emission spectra when excited at 250 nm shows the very characteristic $^5D_0 \rightarrow ^7F_J$ transition of Europium with a stronger peak at 611.2 nm (J=2) typical of the cubic phase sesquioxydes.[38, 39] The other weaker lines are also characteristic of the other transitions toward the $^7F_J$ levels either in the $C_2$ or $S_6$ sites. From this first analysis it can be concluded that the phase is preserved here again. A very slight broadening of the spectra in the case of nanoparticles can also be noticed. More interesting is the completely different emission spectra obtained for the nanoparticles when excited at 350 nm. There, the spectrum only shows large bands for each $^5D_0 \rightarrow ^7F_J$ denoting a stronger disorder. This is generally observed for very small particles because the intra-configurational $4f^6$ of europium are weakly sensitive to the surrounding.[40] The excitation spectra of the 611.2 nm is quite similar for bulk and nanoparticles except that for bulk materials, there is a saturation in absorption which artificially diminishes the spectra below 225 nm due to the optical thickness of the bulk sample. The broad band around 240 nm is due to the charge transfer (CT) from $O^{2-}$ to $Eu^{3+}$. The strong increase in absorption around 225 nm is the consequence of the $Y_2O_3$ band gap fundamental absorption. At longer wavelengths, a series of weak narrow lines corresponds to the intra-configurational $4f^6$ absorption of Europium. In contrast, the excitation spectrum of the emission at 620nm shows some important differences. The $4f^6$ absorptions of Europium remain unchanged, but the CT band is shifted towards longer wavelength and broadened. An additional band around 350 nm also appears, typical of the very small sizes [11]. Finally the shoulder due to the gap is shifted towards shorter wavelengths as expected from a quantum confinement effect.[6, 41] While the low amount of powder produced does not allow a good evaluation of the light yield of photoluminescence, in the case of Y2O3:Eu3+ we can make an indirect estimation of it. Indeed as we mentioned before in the excitation spectra of the bulk sample we observe saturation in absorption below 225nm which was not the case for the nanoparticles. When excited at 250nm (where none of them are saturated), the luminescence of the nanoparticles is 0.75 times that of the bulk. Therefore we conclude that at worst the nanoparticles luminescence is 75% that of the bulk when excited in the charge transfer band.

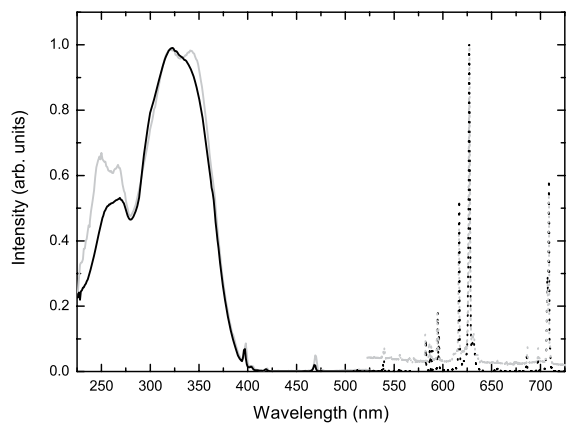

**Figure 10**: Emission (dashed lines) and excitation (continuous lines) luminescence spectra of bulk (black) and nanoparticles (grey) grown by PLAL of Europium doped $Lu_2O_2S$. For the excitation spectra the light was detected at 625nm, for emission spectra the samples were excited at 350nm.

Figure 10 presents excitation and emission spectra for the bulk target of $Lu_2O_2S:Eu^{3+}$ and the nanoparticles obtained by PLAL. For the excitation spectra, the emission was integrated between 626 and 630 nm. For both samples luminescence lines are due to the intra-configurational $^5D_{0-1}\rightarrow{}^7F_J$ transitions of $Eu^{3+}$ with a strongly pronounced peak at 626 nm characteristic of the emission of $Eu^{3+}$ in oxysulfide materials. In the case of the nanoparticles, there is an additional weak broad emission extending all over the visible spectrum. In the excitation spectra sharp lines are observed due to intra-configurational transitions from $^7F_0\rightarrow{}^5D_J$ around 538 nm (J=1) 470 nm (J=2) and 398 nm (J=3). Between 290 and 360 nm there is a broad absorption due to the charge transfer between $S^{2-}$ and $Eu^{3+}$ and below 270 nm another large band attributed to the charge transfer between $O^{2-}$ and $Eu^{3+}$. In the case of the nanoparticles these two CT bands decompose each other into two peaks, probably due to two different sites.

From the spectroscopic results there is no evidence of Europium in $Lu_2O_3$ phase even by searching specifically for it. This is apparently in contradiction with the results from the results of figure 5 and 6. However in the case of the luminescence spectra the solution was lyophilized just after preparation while for the TEM and XRD analysis the solutions were kept for a few weeks after preparation in water and then poured on the TEM grids or lyophilized for XRD. It is therefore highly probable that the oxysulfide nanoparticles underwent some oxidation in the water solution. Indeed the luminescent of the powder was checked again 2 months later and still did not show any $Lu_2O_3:Eu^{3+}$ related luminescence.

**4. Conclusions**

In this paper we have shown how pulsed laser ablation in liquids can be used as a fast method to synthesize complex materials at the nanoscale for four different material families, the sesquioxydes, the oxysulfide, the silicates and the tantalate. This technique can be widely applied to other materials of the same families. Indeed we have been able to synthesize other compositions ($Y_2SiO_5:Ce^{3+}$, $GdTaO_4:Tb^{3+}$, $Gd_{0.4}Y_{2.6}Al_5O_{12}:Ce^{3+}$) which are not presented here

Both phases and stoichiometries are preserved even if the luminescence measurements show the lattice disorder classically observed for the smaller nanoparticles. The size distributions are mono-dispersed and characterized by a mean size of roughly seven nanometers. This last point is particularly interesting because those sizes are big enough to preserve the luminescent properties of the bulk materials, and small enough to find an interest in applications such as bio-labeling or anti-counterfeit marking.

ACKNOWLEDGMENT: We would like to thank Dr. L. Guy for the lyophilization processes and G. Breton, Y. Guillin and S. Orival for technical support. We also would like to thank R. Vera from the center for diffraction Henri Longchambon for invaluable help on the preparation of the samples for XRD.

Facile and rapid synthesis of highly luminescent nanoparticles via Pulsed Laser Ablation in Liquid

SUPPORTING INFORMATION PARAGRAPH:
Energy dispersive X-Ray spectroscopy (EDS) for samples (a) $Lu_3TaO_7:Gd^{3+},Tb^{3+}$, (b) $Gd_2SiO_5:Ce^{3+}$ and (c) $Lu_2O_2S:Eu^{3+}$.

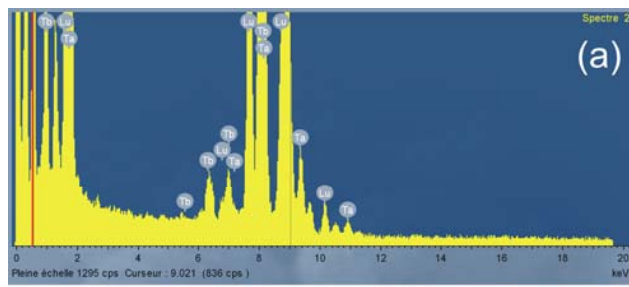
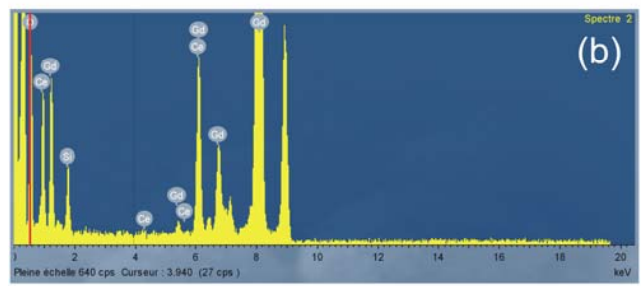
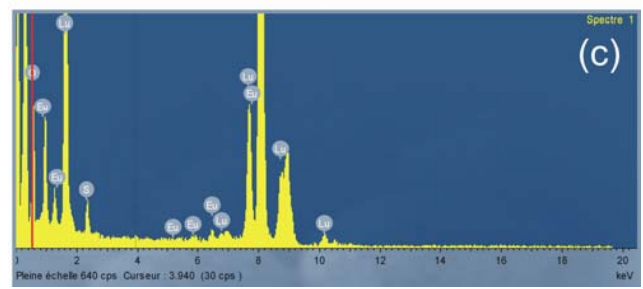

Facile and rapid synthesis of highly luminescent nanoparticles via Pulsed Laser Ablation in Liquid

FIGURE CAPTIONS

**Figure 1**: Scheme of the PLAL experiment. The laser is focused on the target which is rotated. A plasma is generated from which the nanoparticles are formed. On the picture, one can see the impact on the target as well as the plasma plume and its luminescence.

**Figure 2**: Transmission electron microscopy images of four different samples at low magnification: (a) $Lu_3TaO_7$:$Gd^{3+}$,$Tb^{3+}$, (b) $Y_2O_3$:$Eu^{3+}$, (c) $Gd_2SiO_5$:$Ce^{3+}$ and (d) $Lu_2O_2S$:$Eu^{3+}$

**Figure 3**: Examples of image analysis used to get the size distribution in the case of Gd2SiO5:Ce3+. The white circles are guide to the eye giving the average diameter of each individual particle. For higher resolution some particles (those which are well oriented with the electron beam) show diffracting planes (see image d) which helps in the determination of the limit of the particles. (e) Example of size distribution deduced from the analysis of the TEM images. The distribution is well fitted with a log normal distribution with an average size of 8.9nm.

**Figure 4**: High resolution transmission electron microscopy images of 4 different samples: (a) $Lu_3TaO_7$:$Gd^{3+}$,$Tb^{3+}$, (b) $Y_2O_3$:$Eu^{3+}$, (c) $Gd_2SiO_5$:$Ce^{3+}$ and (d) $Lu_2O_2S$:$Eu^{3+}$. In the case of $Gd_2SiO_5$:$Ce^{3+}$ (Image (c)) one can clearly see a nanoparticle for which two families of planes are apparent and which form an angle of 71°.

**Figure 5**: Electron diffractogram for the 4 different samples of figure 4: (a) $Lu_3TaO_7$:$Gd^{3+}$,$Tb^{3+}$, (b) $Y_2O_3$:$Eu^{3+}$, (c) $Gd_2SiO_5$:$Ce^{3+}$ and (d) $Lu_2O_2S$:$Eu^{3+}$. In each case the corresponding diffracting plane families are indicated. In the case of $Gd_2SiO_5$:$Ce^{3+}$ (image (c)) planes -202 and 021 correspond to the same particle and form an angle of 71° as expected. For $Lu_2O_2S$:$Eu^{3+}$ (image (d)) both $Lu_2O_3$ and $Lu_2O_2S$:$Eu^{3+}$ family planes are observed. Those of $Lu_2O_3$ are indicated in the right part of the diffractogram while those of $Lu_2O_2S$:$Eu^{3+}$ are indicated in the left part.

**Figure 6**: X-Ray diffraction for the 4 different samples: (a) $Lu_3TaO_7$ :$Gd^{3+}$,$Tb^{3+}$ (continuous line) compared with JCPDS file n°024-0697 of $Lu_3TaO_7$ (★); (b) $Y_2O_3$:$Eu^{3+}$ (continuous line) compared with JCPDS file n° JCPDS 025-1011 of $Y_{1.9}Eu_{0.1}O_3$ (★); (c) $Gd_2SiO_5$:$Ce^{3+}$ (continuous line) compared with JCPDS file n° 00-040-0287 of $Gd_2SiO_5$ (★) (d) $Lu_2O_2S$:$Eu^{3+}$ (continuous line) compared with JCPDS file n° 026-1445 of $Lu_2O_2S$ (★) and JCPDS file n° 012-0728 of $Lu_2O_3$.(▼)

**Figure 7**: Emission (dashed lines) and excitation (continuous lines) luminescence spectra of bulk (black) and nanoparticles (grey) grown by PLAL of Cerium doped $Gd_2SiO_5$. For the excitation spectra the light was detected at 450nm, for the emission spectra the samples were excited at 290nm.

**Figure 8**: Emission (dashed lines) and excitation (continuous lines) luminescence spectra of bulk (black) and nanoparticles (grey) grown by PLAL of Terbium/Gadolinium doped $Lu_3TaO_7$. For excitation spectra the light was detected at 550nm, for emission spectra the samples were excited at 275nm except for the spectrum with a continuous grey line in the inset that was excited at 325nm.

**Figure 9**: Excitation spectra of Europium doped yttrium oxide nanoparticles grown by PLAL (grey lines) for the emission at 611nm (continuous line) or at 620nm (dashed line) or in the bulk phase (black line). In the inset emission spectra of bulk (black line) excited at 250nm and of the nanoparticles excited at 250nm (continuous line) and 350nm (dashed line).

**Figure 10**: Emission (dashed lines) and excitation (continuous lines) luminescence spectra of bulk (black) and nanoparticles (grey) grown by PLAL of Europium doped $Lu_2O_2S$. For the excitation spectra the light was detected at 625nm, for emission spectra the samples were excited at 350nm.

Facile and rapid synthesis of highly luminescent nanoparticles via Pulsed Laser Ablation in Liquid

TABLES:

Table 1: statistical analysis of the size distribution for samples $Lu_3TaO_7:Gd^{3+},Tb^{3+}$, $Y_2O_3:Eu^{3+}$ and $Gd_2SiO_5:Ce^{3+}$

| samples | Number of particles analyzed | Average diameter (nm) | Standard deviation (nm) | 90% of the particles | 98% of the particles |
|---|---|---|---|---|---|
| $Lu_3TaO_7:Gd^{3+},Tb^{3+}$ | 182 | 6.2 | 2.5 | 4-18nm | 3-22nm |
| $Y_2O_3:Eu^{3+}$ | 124 | 7.0 | 2.8 | 5-20nm | 4-26nm |
| $Gd_2SiO_5:Ce^{3+}$ | 181 | 8.9 | 2.4 | 6-16nm | 4-20nm |

Table 2: Atomic percentage of the elements from the EDS analysis for samples $Lu_3TaO_7:Gd^{3+},Tb^{3+}$, $Gd_2SiO_5:Ce^{3+}$ and $Lu_2O_2S:Eu^{3+}$.

| | | Lu | Gd | Si | O | S | Ta | Eu | Tb | Ce |
|---|---|---|---|---|---|---|---|---|---|---|
| $Lu_3TaO_7:Gd^{3+},Tb^{3+}$ | target | 25 | 0.5 | | 64 | | 9 | | 1.5 | |
| | EDS | 20 ±2 | <1[a] | - | 64 ±6 | - | 13.5 ±1 | - | 2.5 ±1 | - |
| $Gd_2SiO_5:Ce^{3+}$ | target | | 24.75 | 12.5 | 62.5 | | | | | 0.25 |
| | EDS | - | 22 ±1 | 8.5 ±1 | 69.5 ±4 | - | - | - | - | <1[a] |
| $Lu_2O_2S:Eu^{3+}$ | target | 38 | | | 40 | 20 | | 2 | | |
| | EDS | 27.5 ±1 | - | - | 65 ±4 | 6 ±1 | - | 1.5 | - | - |

[a] Less than one percent means not detected. The EDS detection threshold is one percent.

Table 3: Parameters used for the synthesis of the different materials

| Sample | duration | fluence on the target surface |
|---|---|---|
| $Lu_3TaO_7:Gd^{3+},Tb^{3+}$ | 2h | 20.7 GW.cm$^{-2}$ |
| $Y_2O_3:Eu^{3+}$ | 2h15min | 20.7 GW.cm$^{-2}$ |
| $Gd_2SiO_5:Ce^{3+}$ | 2h 10min | 1.67 GW.cm$^{-2}$ |
| $Lu_2O_2S:Eu^{3+}$ | 3h | 1.77 GW.cm$^{-2}$ |